\documentstyle[12pt]{article}
\begin{document}
%\draft
\title {PRIMARY  POSTULATES  OF  THE  STANDARD  MODEL
AS CONSEQUENCES OF THE COMPOSITE  NATURE  OF THE  FUNDAMENTAL
FERMIONS}
\author {Michael A. Ivanov \\
Research Institute of Applied Physical Problems,\\
Belorussian State University, \\
Kurchatov Street 7, 220064, Minsk, USSR \\
Present address: Chair of Physics,\\
Belarus State University of Informatics
and Radioelectronics, \\ 6 P. Brovka Street,  BY 220027, Minsk, Republic of
Belarus. \\ E-mail: ivanovma@gw.bsuir.unibel.by.}
\date{July 1, 2002} \maketitle
\begin{abstract}
A field model of two-component fermions is described,  the
consequences of which coincide in the main with primary postulates
of the standard model. Such a model can be constructed for 4
generations at the minimum.  Peculiarities of the relative
coordinate space, determining in general an internal symmetry
group, are considered. Analogues of the Higgs fields appear in the
model naturally after transition to the Grassmannian extra
coordinates.
\end{abstract}

%Reprint, LaTeX, 20 pages. Few references and remarks are added,
%some found mistakes are corrected
%Nuovo Cimento, 105A (1992) 77-89.

\section[1]{Introduction }
The standard model (see \cite{1}), the result of collective work
of many physicists which has not today any non-accordance with
phenomenology, contains a number of postulates both very
complicated and having weak logical ties among themselves. For
example, there are postulates about the internal symmetry group,
the existence of a number of generations, transformation of
multiplets in different generations by the interwoven
representations of the symmetry group and the existence of the
Higgs scalar fields. This complication of postulates is conserved
also in the Grand Unification theories \cite{2,3,4}, based on
different gauge groups. Spontaneous breaking of a symmetry to the
standard model symmetry in such theories is one way to make a
theory agree with experimental data. Another way of concordance we
see in the theories of the Kaluza-Klein type \cite{5,6,7,8} when
by compactification of the additional space dimensions one can
introduce manifolds with the symmetry group that occurs in the
standard model. These ways are similar while in GUTs one talks
about a symmetry group which will be reduced, but in the
Kaluza-Klein theories one builds at first a geometrical
construction and then only uses its symmetry. In the mechanisms of
spontaneous compactification of additional dimensions, the
auxiliary fields are used too \cite{9}, as in GUTs similar fields
are used for breaking of a symmetry. In both the standard model
and GUTs one does not raise a question about an internal symmetry
nature, and in the Kaluza-Klein theories about the physical sense
of the additional dimensions.
\par
In the supersymmetrical models an internal symmetry group arises
as result of an equal state of the spinor coordinates
\cite{10,11}. Usually in such models a special stress is laid on
the symmetry between fermions and bosons, but it is typical only
for the models on the basis of superspaces with torsion when the
additional coordinates are transformed by the spinor
representations of the Lorentz group of the basic space. For
conservation of an internal $SU(N)-$symmetry an availability of
superspace torsion is not obligatory.
\par
In the $E_{8}\times E_{8}-$superstring theory, which is the most
likely candidate to the role of the realistic Kaluza-Klein theory,
a spectrum of the massless four-dimensional fields, appearing as
exitations upon the compactified main state of the string,
contains fields of four generations of chiral fermions \cite{12}.
\par
In this paper a model of the composite fermions - to be more
exact, a model of two-component fermions - \cite{13,14,15} is
described in English; this one is a theory of the Kaluza-Klein
type but with the peculiarity that the internal space in it must
be discrete. It turns out that consequences of this model either
literally repeat the primary postulates of the standard model - a
global symmetry group in a generation is $SU(3)_{c} \times
SU(2)_{l} \times U(1),$ multiplets in different generations must
be transformed by the interwoven representations of the group - or
differ from corresponding postulates not essentially - in
particular, the composite fermions make up 4 generations in the
model, a permissible multiplet of the model contains all states of
the standard model multiplet but may contain another SU(2)-singlet
in the lepton sector \footnote{It gives us a possibility to
introduce neutrino masses that is important now - after
observation of evidence that these masses are non-zero}. After
transition to the Grassmanian variables as additional coordinates,
the $SU(2)-$doublets appear in the model naturally, which are the
Lorentz-scalar ones and which may be identified with the Higgs
fields. Such a coincidence of the composite fermion model
consequences with the postulates of the standard model, which is
well verified by experience, can be interpreted as an indirect
confirmation of the compositeness of the fundamental fermions and
this is deserving great attention. On the basis of the simple
physical idea about the composite nature of fermions it is
possible to logically tie up the main postulates of the standard
model among themselves, and it takes away at least one problem -
the problem of generations: in the composite fermions model a set
of generations appears so naturally as a particle and an
antiparticle in the Dirac theory and it is impossible to build a
similar model for one generation. In such a context the standard
model comes forward as a way for the non-evident description of
non-local interactions, the problem whose decision one attempted
to find by constructing non- local field theories in the 4-space
\cite{16}.
\par
The model has a clear physical interpretation of the additional
coordinates - these are coordinates of the relative location of
the constituents of a composite system.
\par
Usually in the composite particle models \cite{17,18,19} it is
assumed that its constituents have definite properties on
transformations of the 4-space coordinates: there are fermions -
charged and neutral rishons in \cite{18}, spinor and scalar
subquarks in \cite{19}. In contrast to this, in our model
constituents of a composite system have not definite properties
with respect to transformations of coordinates of its centre of
inertia, and the relative coordinates do not form the Minkowski
space.
\par
Constants of the length dimension appear in the model naturally -
four in the lepton sector and one in the quark sector, but only an
analysis of the global symmetries of the model does not permit to
evaluate their magnitudes and to state any relations among them,
and also to compare ones with a fundamental length of other models
\cite{20}. It is clear only that all of them cannot be equal to
zero - it is necessary to choose the symmetry group $SU(3)_{c}
\times  SU(2)_{l} \times  U(1).$ \par It is shown in \cite{15} that
this model of the composite fermions may considered in the
$(N=2)-$superspace without torsion, with the additional
coordinates which are independent of the main ones. A number and a
character of ties, choosing a set of supermanifolds, determine an
internal symmetry group, while in the supersymmetrical models this
one is determined by an extension degree N.

\section[2]{Some algebraic remarks }

The composite fermion model under consideration is based on one
generalization of the Dirac procedure \cite{21} of construction of
a linear equation for a multicomponent function, and it is
appropriate to consider an algebraic side of things. The Dirac
procedure of transition from an equation $E^{2}=\bar{p}^{2} + m^{2}$
to a linear equation for a new multicomponent function $ \psi_{1}
=(\psi ,... ,\psi_{4})^{T}$ can be devided into two stages: an
algebraic stage of linearization of this equation, which is not
connected with the introduction of new postulates:
\begin{equation}
E\psi =\beta m\psi +\alpha_{k}p_{k}\psi,
\end{equation}
where $\beta$ and $\alpha_{k}$ are the $4\times 4-$matrices, $E,
m, p_{k}$ do not change, and another stage of transition to the
quantum description, connected with the introduction of a
postulate about replacement of $E$ and $p_{k}$  by operators of
motion in the space. Here we consider namely the first stage;
further it will be used to connect vectorial and spinor additional
coordinates. \par This Dirac procedure can be generalized to
\begin{equation}
E=\sum^{N}_{i=1}( p_{1}^{r} + p_{2}^{r}  +...+p_{n}^{r})^{1/k},
\end{equation}
with different $N, n, k, r;$ in this paper we deal with $N=k=r=2,
n=4,$ i.e. with an equation for energy of two free particles
\cite{13}.
\par
Especially the case of $N=1, k=r$ must be picked out, i.e.
equations
\begin{equation}
E^{r}= p_{1}^{r} + p_{2}^{r}  +...+p_{n}^{r}.
\end{equation}
Probably, eqs. (3) with $r>2$ have not a particular relation with
the physical problems though such possibility was considered by
Jamaleev \cite{22}; but these are important for the number theory,
and I confine myself to formulate a theorem, connected with those
whose constructive proof for prime $r$ has been produced in
\cite{23}. This theorem is interesting by the circumstance that in
it for an algebraic problem "a wave function", well known to
physicists, is introduced - it makes one think.
\newtheorem{t1}{Theorem \cite{23}} \nonumber
\begin{t1}
With eq. (3), $n \geq 2,$ a linear matrix equation
\begin{equation}
E\psi = p_{1}A_{1}\psi+p_{2}A_{2}\psi+...+p_{n}A_{n}\psi,
\end{equation}
can be juxtaposed, where $A_{i}$ are the quadratic matrices of order
$mn$, for which the condition holds
\begin{equation}
(p_{1}A_{1}+p_{2}A_{2}+...+p_{n}A_{n})^{r}= p_{1}^{r} + p_{2}^{r}
+...+p_{n}^{r},
\end{equation}
$\psi$ is the column-vector $(\psi_{1},...,\psi_{mn})^{T}$; if (3)
is carried out, a determinant of the system of equations (4) with
respect to unknown quantities $\psi_{i}$ is equal to zero.
\end{t1}
The proof of this theorem is based on the study of the properties
of algebraic objects, called in \cite{23} generalized
anticommutators: $$(ABC...)_{+}\equiv \sum_{P} ABC...,$$ where
summing up has been made on all different permutations of matrices
$A,B,C,....$\par Let us consider the simplest example of eq. (4)
to show that "a wave function" in number theory problems - only
without such a name - is well known for mathematicians. For the
Diophantine equation  $z^{2} = x^{2}+y^{2}$ the parametrization of
its solutions: $x = 2\psi_{1}\psi_{2},\ y =\psi_{1}^{2} -
\psi_{2}^{2},\ z = \psi_{1}^{2} + \psi_{2}^{2}$ is known; but the
solutions of then equation $$z\psi =\sigma_{1} x\psi +\sigma_{3}y
\psi,$$ where $\sigma_{i}$ are the Pauli matrices, are just of
this kind, if $\psi= (\psi_{1} ,\psi_{2})^{T}.$ \par By
linearization of an equation
\begin{equation}
E  = (m_{1}^{2}+\bar{p}_{1}^{2})^{1/2} + (m_{2}^{2}+\bar{p}
 _{2}^{2})^{1/2}
\end{equation}
considered below, defining relations of an algebra of introduced
matrices contain both commutators and anticommutators.

\section[3]{Description of the two-component composite system }

To make the physical idea clear, lying on the basis of the
composite fermion model, before formulation of its postulates we
consider the description of a composite system with the help of a
wave function in the $8-$space \cite{13}. With eq. (6) we
juxtapose a linear equation $(c = h = 1)$
\begin{equation}
E\psi=(\beta^{1}m_{1}+\alpha^{1k}p_{1k}+\beta^{2}m_{2}+
\alpha^{2k}p_{2k})\psi
\end{equation}
and substitute $E,\ p_{ik}$ by operators of system energy and
constituent momenta in it. To obey the conformity principle,
matrices $\beta^{i}, \ \alpha^{ik}$ should have the dimension
$16\times 16$ and should satisfy the conditions:
\begin{eqnarray}
\{\alpha^{ik},\alpha^{il}\}_{+}=2\delta^{kl},
\{\alpha^{il},\beta^{i} \}=0, \nonumber\\ \left[
\beta^{i},\beta^{j} \right]_{-}=\left[
\alpha^{ik},\beta^{j}\right]= \left[
\alpha^{ik},\alpha^{jl}\right]=0, \nonumber \\ i\neq j, \
\beta^{i2}=I^{16} \
\end{eqnarray}
(there is not summation on $i$ in the last line). The form of eq.
(7) coincides with that of the Bethe-Salpeter equation
\cite{24,25}, which is used for the description of such composite
systems as mesons. \par Further interpretation of eq. (7) depends
essentially on geometrical properties, which are provided by the
eight-dimensional coordinate space of a model: if one suggests
that coordinates of every constituent $x_{i}^{\mu}$ belong to the
Minkowski space $R^{4}_{1,3}$ we get a description of a system of
two fermions. Another possibility, considered in \cite{13,14},
leads to the composite fermions model: it is assumed in this one
that only coordinates of a centre of inertia $x^{\mu}$ belong to
$R^{4}_{1,3}$ while about internal coordinates (the ones of the
relative location) it is assumed only, that these are transformed
independently $x^{\mu}.$ Besides that, additional restrictions are
taken on a wave function which do not contain evidently
derivatives with respect to time, to be like the condition on a
wave function for particles with spin $3/2,$ when field equations
are written in the Rarita-Schwinger form \cite{26}. \par Let us
define four-momentum operators $$p_{\mu}=i\partial/\partial
x^{\mu},\ \pi_{\mu}=i\partial/\partial y^{\mu},\
p_{i\mu}=i\partial/\partial x^{\mu}_{i},\ i=1,2,\ \mu=0,1,2,3,$$
and designate corresponding coordinate spaces by $X,\ Y,\ X_{1},\
X_{2}.$ We deal with two $8-$spaces, connected among themselves by
a linear transformation. If $z=(x,y),\ z_{12}=(x_{1},x_{2}),$ then
these are $Z=\{z\}$ and $Z_{12} =\{z_{12}\}.$ In eq. (7) operators
of motion in both $Z$ and $Z_{12}$ enter; as was shown in
\cite{13}, if $p_{\mu}\psi =p_{1\mu}\psi+p_{2\mu}\psi,\
\pi_{\mu}\psi =p_{1\mu}\psi-p_{2\mu}\psi,$ the additional
condition on $\psi \ (k=1,2,3)$
\begin{equation}
\alpha^{1k}p_{1k}\psi +\alpha^{2k}p_{2k}\psi =A^{k}p_{k}\psi,
\end{equation}
with some matrices $A^{k}$ permits to split (7) into four Dirac
equations for some four components of $\psi.$ If one considers
that every constituent of a composite system moves according to
the equation
\begin{equation}
p_{i0}\psi=\alpha^{ik}p_{ik}\psi + \beta^{i} m_{i}\psi,
\end{equation}
one gets that a system motion is described by equations
\begin{equation}
E\psi= (\beta^{1} m_{1}+\beta^{2} m_{2})\psi+\bar{A} \cdot
\bar{p}\psi,
\end{equation}
\begin{equation}
\pi_{0}\psi= (\beta^{1} m_{1}-\beta^{2} m_{2})\psi+\bar{A} \cdot
\bar{\pi}\psi.
\end{equation}
We have equivalent descriptions of the system, using only eqs.
$(10)-(12)$ or $(7)-(10)$. Later on eqs. $(10)-(12)$ with
$m_{1}=m_{2}=0$ are taken for the basis. In such a manner, at
present we make certain that the description of a composite system
can be done such that it will correspond for $m_{1}=m_{2}=0$ to
four "generations" of massless fermions. What symmetries has this
model? But before we should formulate clearly primary postulates
of the model and write its equations without picking out operators
of energy \cite{14}.
\section[4]{Primary postulates}
Let us introduce the following postulates: \par
%\newtheorem{t2}{}\nonumber
%\begin{t2}
{\it a) a fermion is a composite two-component system; \par b) the
coordinate space is an eight-dimensional and real one, the
coordinates of a centre of inertia $x^{\mu}$ set up the Minkowski
space $R_{1,3}^{4};$\par c) internal coordinates $y^{\mu}$ are
transformed independently of $x^{\mu}$ (compare with \cite{27}).}
%\end{t2}
\par Namely the latter of them leads in the model to the consequence:
the internal coordinate space is not the Minkowski one. To speak
about description with the help of this model of four generations
of fermions in a space $X,$ it is necessary to have a possibility
to average on coordinates $y$ the function $\psi(x,y),$
introducing in (11) or its analogue, with conservation of the
equation form; but the independence of $y$ on $x$ is necessary for
it.
\par
It is assumed that the physical space-time coincides with $X.$

\section[5]{Field equations of the two-component fermions}

Taking into account our designations, the main equations for the
$16-$component wave function of massless, non-charged, free
fermions can be written as four Dirac-Kahler equations (see
\cite{28}):
\begin{equation}
\Gamma_{1}^{\mu}p_{\mu}\psi(x,y)=0,
\end{equation}
\begin{equation}
\Gamma_{1}^{\mu}\pi_{\mu}\psi(x,y)=0,
\end{equation}
\begin{equation}
\Gamma_{2}^{\mu}p_{1\mu}\psi(x_{1},x_{2})=0,
\end{equation}
\begin{equation}
\Gamma_{3}^{\mu}p_{2\mu}\psi(x_{1},x_{2})=0,
\end{equation}
which are equivalent to $(10)-(12),$ if matrices
$\Gamma_{a}^{\mu},\ a=1,2,3,$ obey conditions (compare with (8)):
\begin{equation}
\{\Gamma_{1}^{\mu},\Gamma_{1}^{\nu}\}=
\{\Gamma_{2}^{\mu},\Gamma_{2}^{\nu}\}=
\{\Gamma_{3}^{'\mu},\Gamma_{3}^{'\nu}\}=2g^{\mu\nu}, \
\{\Gamma_{2}^{\mu},\Gamma_{3}^{'\nu}\}=0,
\end{equation}
where $g^{\mu\nu}=diag(1,-1,-1,-1),\ \Gamma_{3}^{'\mu}\equiv K
\Gamma_{3}^{'\mu};\ K=\sigma^{3}\times I^{8}$ for the
representation of $\Gamma_{a}^{\mu}$ from \cite{14}.
\par
As in the Dirac theory, eqs. $(13)-(16)$ must be considered also
as postulates of the model. These contain non-evidently condition
(9), connected not only with coordinate spaces, but also with a
wave function. \par After averaging of $\psi(x,y)$ with respect to
$y$ in (13): $\tilde{\psi}(x)=<\psi(x,y)>_{y}$ by taking into
account the postulate $c),$ we see that the model really
corresponds to four generations of fermions, being described by
$\tilde{\psi}(x).$ Though (15),(16) have the same structure, one
cannot say that the two-component system is composed by two
fermions - below it will be shown that, for taken axioms,
permissible transformations of coordinates $x_{1},x_{2}$ do not
belong to the Lorentz group or the Poincare one.

\section[6]{A kind of permissible transformations of coordinates}

Coordinates of two eight-dimensional spaces $Z$ and $Z_{12}$ have
been connected by the linear transformation with a matrix
$\lambda$ (compare with \cite{17}):
\begin{equation}
z^{\alpha}=\lambda^{\alpha}_{\beta}z^{\beta}_{12},\
\alpha,\beta=1,...,8,
\end{equation}
where $\lambda=(\sigma^{1}+\sigma^{3})\times I^{4}/2,\
\lambda^{-1}=2\lambda,\  |\lambda|=-2^{-4}.$ Accordingly operators
$\partial_{\alpha}=i\partial/\partial z^{\alpha}$ and
$D_{\alpha}=i\partial/\partial z^{\alpha}_{12}$ (i.e.
$\partial=(p,\pi),\  D=(p_{1},p_{2})$) have been connected
by the formula
$$D_{\alpha}=\lambda^{\alpha}_{\beta}\partial_{\beta}.$$
\par
In accordance with $c),$ matrices of permissible transformations
of coordinates of $Z$ should have the block-diagonal kind:
\begin{equation}
\Lambda=diag(\Lambda_{x},\Lambda_{y}),
\end{equation}
where $\Lambda_{x}$ and $\Lambda_{y}$ are matrices of
transformations of $x$ and $y.$ The transformation $z_{12}
\rightarrow Lz_{12}$ corresponds to the one $z \rightarrow \Lambda
z $ with
\begin{equation}
L=\lambda^{-1}\Lambda\lambda=\left[I^{2}\times(\Lambda_{x}+\Lambda_{y})
+ \sigma^{1}\times(\Lambda_{x}-\Lambda_{y})\right]/2.
\end{equation}
It is natural to interpret transformations
$\Lambda^{'}=diag(\Lambda_{x},I^{4}),$ where $\Lambda_{x}$ is the
Lorentz transformation, as the "coordinate" ones of the physical
space-time, and these of the kind
$\Lambda^{''}=diag(I^{4},\Lambda_{y})$ - as the "internal
symmetry" ones. There exists every
$\Lambda=\Lambda^{'}\Lambda^{''}.$
\par
A matrix $L$ is not diagonal for $\Lambda_{x}\neq \Lambda_{y}.$

\section[7]{A global symmetry group of the model}

An internal space turns out to be discrete for extra coordinates
$y$ of the vectorial kind in the model, in contrast to the
situation in others Kaluza-Klein theories. In such a case a
continuous group may be a symmetry group of the model, if an
additional postulate about conservation of a norm for a set of
model solutions is introduced \cite{14}. Probably, the model may
be supplemented  - for example, by introduction of some
"observation space", differing from the "existence space" of
fermions - in such a manner, that one will manage to find a proof
of this postulate. It is probable also that the restriction
$\pi_{\mu}\psi(z)=0,$ taken below from the model covariance
condition, may be replaced by some softer one.
\par
A group $SU(n)$ is a symmetry one for a field model if \par
%\begin{t3}
{\it 1) a model permits $n$ different solutions
$\psi_{A},\psi_{B},...;$\par 2) a direct transition from any
$\psi_{A}$ to any $\psi_{B}$ is possible;\par 3) any linear
composition of these solutions is a model solution too;\par 4) for
$\varphi^{T} \equiv (\psi_{A},\psi_{B},...)$ the norm
$\varphi^{\dag}\varphi$ is conserved.}
%\end{t3} \nonumber
\par Three first signs of the presence of a group $SU(n)$ can be
ascertained in this model and the fourth one may be introduced as
an additional postulate. Such a postulate is much simpler than the
choice of the definite group in the standard model.
\par
It was shown in \cite{13} that model $(13)-(16)$ permits eight
different solutions, which are $G_{A}\psi(x,y),$ if $\psi(x,y)$ is
some solution. These solutions are broken up by the sign 2) on
sets of $1,2,3$ elements, thus taking condition 4) of conservation
of the norm $\varphi^{\dag}\varphi$ for the sets as postulate, one
has that $SU(3)_{c}\times SU(2)_{l}\times U(1)$ appears as the
global symmetry group of the model. Matrices $G_{A},\ A=1,...,8,$
set up a representation of the discrete group $D_{4}$ \cite{29}.
The discrete symmetry groups are used in field theories - for
example, see \cite{30}. Indeed, as one may check, eight matrices
$G_{A}$ exist for eq. (13), such that
$[G_{A},\Gamma_{1}^{\mu}]=0.$ The representation has been given
for them in \cite{14}. The multiplication law of these corresponds
to the one of $D_{4}.$ One of the ideas of the composite fermions
model consists in the fact that eight solutions of eq. (13),(14)
of the kind $\psi_{A}(x,y) =G_{A}\psi(x,y)$ have been assumed to
be induced by transformations $\Lambda_{yA}$ of a space $Y,$ also
setting up a representation of $D_{4}.$ It leads to the fact that
a given algebraic structure of field equations puts hard
restrictions on the geometrical properties of a space of internal
coordinates. An additional demand of covariance of these equations
leads to a set of matrices $\Lambda_{yA}$ with imaginary elements,
and in this connection the postulate b) can be conserved only at
the price of fixing of a definite topology for every $A.$ If
$\Lambda_{A}= diag(I^{4},\Lambda_{yA}),$ a transformation
$Z\rightarrow \Lambda_{A} Z$ would be accompanied by
$\psi(z)\rightarrow \psi_{A}(z),$ and a condition of covariance of
(14) will be
\begin{equation}
\Gamma^{\mu}_{1\pm}= M_{A}\Gamma^{\nu}_{1\pm}\Lambda_{yA\nu}^{\mu}
G_{A},
\end{equation}
where $\Gamma^{\mu}_{a\pm}\equiv \Gamma^{\mu}_{a}(1 \pm
\Gamma^{5}), \
\Gamma^{5}=i\Gamma^{0}_{1}\Gamma^{1}_{1}\Gamma^{2}_{1}\Gamma^{3}_{1},\
M_{A}$ is some matrix. It appears that, if $\Lambda_{yA+}$
transforms coordinates in an equation for right components
$\psi_{R}\equiv (1+ \Gamma^{5})\psi /2,$ and $\Lambda_{yA-}$ for
left ones $\psi_{L}\equiv (1- \Gamma^{5})\psi /2,$ then condition
$\Lambda_{yA}=\Lambda_{yA+}=\Lambda_{yA-}$ makes only the trivial
solution of (21) possible: $\Lambda_{yA}=aI^{4}$ \cite{14}. \par
Removing this restriction, i.e. considering $\Lambda_{yA+}\neq
\Lambda_{yA-},$ we obtain from (21) (see \cite{14}) that any
matrix $\Lambda_{yA\pm}$ must have the structure
\begin{equation}
\Lambda_{yA+}=aI^{4}+bB_{1}+cB_{2}+dB_{3},
\Lambda_{yA-}=a^{'}I^{4}+b^{'}B_{1}^{*}+c^{'}B_{2}^{*}+d^{'}B_{3}^{*},
\end{equation}
where $a,b,c,d$ and $a^{'},b^{'},c^{'},d^{'}$ are some numbers,
$*$ is complex conjugation, $B_{1},\ B_{2},\ B_{3}$ are the
matrices $$B_{1}\equiv r^{1}\times \sigma^{1} - r^{3}\times
\sigma^{2},\\
B_{2}\equiv \sigma^{1}\times r^{1} +
\sigma^{2}\times r^{3},\\
B_{3}\equiv r^{2}\times r^{2}+
r^{4}\times r^{4}+ ir^{4}\times r^{4} -ir^{4}\times r^{2},$$ and
matrices $r^{a}$ have the form
\begin{displaymath}
r^{1}=\left(\matrix {1 & 0\cr 0 & 0\cr} \right) ,\ r^{2}=\left(\matrix
{0 & 1\cr 0 & 0\cr} \right),\ r^{3}=\left(\matrix {0 & 0\cr 0 & 1\cr}
\right),\ r^{4}=\left(\matrix {0 & 0\cr 1 & 0\cr} \right).
\end{displaymath}
\par
The following matrices, given upon the field $C,$ set up
representations of $D_{4}\ (B_{k}^{2}=I^{4}):$
\begin{equation}
\{\Lambda_{yA+}\}=\{I^{4},-iB_{1},-B_{2},B_{3};-I^{4},iB_{1},B_{2},-B_{3}
\},\  \Lambda_{yA-}=\Lambda_{yA+}^{*}.
\end{equation}
This fact can be made to agree with postulate b) about the real
type of $Y$ by introduction of restrictions on values of $y^{\mu}$
for every $A.$ If real $y^{\mu}$ corresponds to the case $A=1,$
then the representation (23) does not contradict postulate b) for
the following restrictions on $y^{\mu}:\ y^{0}=y^{1}=0,\ A=2,6;\
y^{1}=y^{3}=0,\ A=3,7;\ y^{1}=y^{2}=0,\ A=4,8,$ i.e. the internal
space for all three solutions with $A= 2,3,4$ and $A=6,7,8$ must
consist of separate two-dimensional planes. For all these 6
solutions (states of field) $y^{1}=0,$ i.e. there is reduction to
the $3-$dimensional pseudo-Euclidean space with the topology:
$y^{0}=0\bigvee y^{2}=0\bigvee y^{3}=0,$ in which its own plane
corresponds to every pair of solutions.
\par
The demand of covariance and eqs. (15),(16) leads, as we shall
see, to further restrictions on internal coordinates - it turns
out that $y^{\mu}$  must be constant, though these may differ for
various $A.$ Let us consider transformations of (15),(16), which
accompany those of $\Lambda$ in $Z.$ The transformation
$\Lambda^{'}\Lambda_{A}$ induces $z_{12} \rightarrow L_{A}z_{12}.$
We designate $\Gamma_{2}^{\alpha}\equiv \{\Gamma_{2}^{\mu}
;0\}^{\alpha},\ \Gamma_{3}^{\alpha}\equiv \{0;\Gamma_{3}^{\mu}
\}^{\alpha},\ \alpha=1,... ,8;$ then eqs. (15),(16) after
transformation are
\begin{equation}
\tilde{\Gamma}_{a}^{\alpha}D^{'}_{\alpha}\psi^{'}(z^{'}_{12}) =
0,\ \alpha=2,3,
\end{equation}
where $D^{'}_{\alpha}=L_{A\alpha}^{\beta}D_{\beta},\
\psi^{'}(z^{'}_{12})= V_{A}\psi(z_{12}),\ V_{A}$ are matrices,
also setting up a representation of $D_{4}$ (this is given in
\cite{14}), and the covariance condition is
\begin{equation}
M_{A}\tilde{\Gamma}_{a}^{\alpha}L^{\beta}_{A\alpha}V_{A}=\Gamma_{a}^{\beta},\
\alpha=2,3.
\end{equation}
It results from (20) that, if $\Lambda_{x}\neq \pm \Lambda_{yA},\
\tilde{\Gamma}_{a}^{\alpha}$ may contain more then 4 nonzero
matrices. A conservation of postulate c) and a covariance of eqs.
(15),(16) can be secured for some $M_{A},\ V_{A},$ if the
condition is carried out
\begin{equation}
p_{1\mu}\psi(z_{12}) =p_{2\mu}\psi(z_{12}).
\end{equation}
In such a case $\pi_{\mu}\psi(z)=0,$ i.e.
$\psi(x,y)=\psi(x,y_{A}^{\mu}),$ where $y_{A}^{\mu}$ are
constants, which may differ for various $A.$ If one assumes that
transitions are possible between all states $\psi_{A}(z),$ then
from restrictions on $y^{\mu}$ for $A=2,3,4;\ 6,7,8$ made above,
it should follow $y_{A}^{\mu}=0 \forall A,$ i.e. the theory should
be local with the symmetry group $SU(8).$
\par
A conservation of a non-local type of the theory is provided by
isolation of field states with $A=1,5$ from the rest; as a result
one has the symmetry group $SU(3)_{c}\times SU(2)_{l}.$ One gets
up such isolation by means of the choice of values of
$y_{A}^{\mu}.$ Then states with $A \neq1,5$ may be interpreted as
the colour states of field since their symmetry group will be
$SU(3).$ Coordinates of the colour space $y_{A}^{\mu},\ A=2,3,4;\
6,7,8,$ can be chosen in such a manner, that matrices of their
transformations $T_{b}$ set up a discrete group, which is
isomorphic to $D_{3}:\ \{T_{b} \mid b=1,...,6\},$ with
$T_{1}=I^{4},\ T_{2}=T_{3}^{-1},\ T_{2}^{3}=T_{4}^{2}=T_{5}^{2}=
T_{6}^{2}=I^{4}.$ A set of $y_{A}^{\mu}$ for such $A$ turns
reserved, i.e. transformations $T_{b}$ do not lead to appearance
of new values of $y_{A}^{\mu},$ only for the condition
\begin{equation}
y_{A}^{\mu}=0 \bigvee y_{A}^{\mu}=\pm l,\ A=2,3,4;\ 6,7,8;
\end{equation}
where $l$ is a constant with dimension of length. Thus all quark
sector of the model has been characterized by one length $l.$ This
condition does not affect the values of $y_{A}^{\mu}$ for $A=1,5,$
so as there are five constants with such dimension (it is
$y_{1}^{\mu}=-y_{5}^{\mu})$ in the model, and four of them concern
the lepton sector. There are not any restrictions, that all these
may be reduced to zero or $\pm l,$ but there are not any grounds
for such conclusion at the global symmetry level.
\par
An action of transformations $T_{b}$ induces transitions
$\psi_{A}(z_{12}) \rightarrow \psi_{A}(z_{12}),$ and it turns out
to be impossible to construct a representation of $D_{3},$ the
matrices of which transform states $\psi(z_{12})$ separately. But
a representation of $D_{3}$ exists, the matrices of which permute
components of triplets $\left(\psi_{A},\ \psi_{A^{'}},\
\psi_{A^{''}} \right),$ where $(A,A^{'},A^{''})=(2,3,4)$ or
$(6,7,8).$ The kind of matrices $S_{b},\ b=1,...,6,$ of this
representation may be easy determined by action of $T_{b}$ on the
set of $y_{A}^{\mu}$ for $A=2,3,4.$ With an additional demand of
conservation of a norm $\varphi^{\dagger} \varphi,\ SU(3)_{c}$
will be the symmetry group of these states.
\par
The $SU(2)$ symmetry is connected with the transformation
$(x,y)\rightarrow (x,-y),$ and the chirality of it has been caused
by the algebraic structure of eqs. (15),(16).
\par
The replacement $y \rightarrow -y$ leads to the permutation
$\left(x_{1},x_{2}\right) \rightarrow \left(x_{2},x_{1}\right),$
so that eqs. (15) and (16) must turn into one another. But the
construction of matrices $\Gamma_{2}^{\mu}$ and $\Gamma_{3}^{\mu}$
permits them to make a transition $(15) \leftrightarrow (16)$ by
means of transformation $\psi(z_{12}) \rightarrow V \psi(z_{12}),$
with some matrix $V$ only for $\psi_{L}$ or $\psi_{R}.$ So, a
matrix $V$ exists with the property
\begin{equation}
\Gamma_{3-}^{\mu}V =V\Gamma_{2-}^{\mu},
\end{equation}
but in this case
\begin{equation}
\Gamma_{3+}^{\mu}V \neq V\Gamma_{2+}^{\mu},
\end{equation}
because components  $\psi_{L}$ and $\psi_{R}$ should be
transformed differently by $y \rightarrow -y.$ Eqs. (15) and (16)
will be covariant, if
\begin{equation}
\psi_{R}(z_{12}) \rightarrow \psi_{R}(z_{12}),\ \psi_{L}(z_{12})
\rightarrow V \psi_{L}(z_{12}),
\end{equation}
and for covariance of (13),(14), by taking into account the
correspondence of matrices $V_{5}=V$ and $G_{5}$ in
representations of $D_{4},$ it is enough
\begin{equation}
\psi_{R}(z) \rightarrow \psi_{R}(z),\ \psi_{L}(z) \rightarrow
G_{5} \psi_{L}(z).
\end{equation}
Thus, for $(x,y)\rightarrow (x,-y)$ solutions $\psi_{R}(z)$ and
$G_{5} \psi_{R}(z)$ are singlets, and $\psi_{L}(z)$ and $G_{5}
\psi_{L}$ are components of a doublet of $SU(2),$ and this is true
also for the colour states with $A=2,3,4;\ 6,7,8.$ Equations
(13)-(16) are invariant also with respect to a global $U(1).$
Because of this, $SU(3)_{c}\times SU(2)_{l}\times U(1)$ will be
the global symmetry group of the model. The structure of a
permissible multiplet of the group differs from this one, taken in
the standard model, only on a possibility of appearance of another
$SU(2)_{r}-$singlet for states with $A=1,5,$ i.e. in the lepton
sector, that may provide a non-zero mass of neutrino.

\section[8]{Covariance of the model under the global Lorentz
transformations} If $\Lambda=(\Lambda_{x},I^{4}),\
\theta_{\mu\nu}$ are parameters of the Lorentz transformation,
$G_{x}=\exp(-i\sigma^{\mu\nu} \theta_{\mu\nu}/4),$ where
$\sigma^{\mu\nu}=[\Gamma^{\mu}_{1},\Gamma^{\nu}_{1} ],$ then a
transformation
\begin{equation}
\psi(z) \rightarrow   G_{x}\psi(z)
\end{equation}
keeps (13) covariant, being $[G_{x},G_{A}] =0.$ Equations
(15),(16) by condition (26) will be covariant, if
\begin{equation}
\psi(z_{12}) \rightarrow   S_{x}\psi(z_{12}),
\end{equation}
where $S_{x}=\exp(-i(\sigma^{\mu\nu}_{2}+\sigma^{\mu\nu}_{3})
\theta_{\mu\nu}/4),\ \sigma^{\mu\nu}_{a}
=[\Gamma^{'\mu}_{a},\Gamma^{'\nu}_{a}],\ a=2,3,\ \Gamma^{'\mu}_{2}
=\Gamma^{\mu}_{2}.$
\par
In the $4-$dimensional case transformation (33) would correspond
to the field states of spin $0,1$ or $1/2$ \cite{28}, but in the
$8-$dimensional case such an interpretation has not any sense. In
the $4-$space-time, states of field $\tilde{\psi}(x),$ describing
four generations of fermions, will be observable.

\section[9]{Introduction of the Grassmannian extra coordinates}

It is shown in \cite{15} that transition from internal coordinates
of the vectorial type $y^{\mu}$ to the Grassmannian ones
$\chi_{a},\bar{\chi}_{a},\ a=1,2,3,4,$ leads to the appearance of
the effective four-dimensional fields in the model's Lagrangian,
which are $SU(2)-$doublets and analogs of the Higgs fields and may
play their part in the mass splitting mechanism.
\par
We show here, following \cite{15}, that one may consider the model
described above as a model of field in the $(N=2)-$superspace
without torsion, with independent additional coordinates, in which
with the help of ties a set of supermanifolds has been picked out,
and one interprets mixing of solutions, to be set on these
supermanifolds, as an internal symmetry. For this, with lack of
torsion there is not any symmetry between bosons and fermions.
Another difference from the supersymmetrical models is that an
internal symmetry group is not defined now by an extension degree
$N,$ but it is given by a character of ties and the supermanifolds
number, which are picked out. An extension with $N=2$ is necessary
only to provide an arbitrary "norm" $y^{\mu}y_{\mu}$ of relative
coordinates - as it is shown in \cite{15}, the $(N=1)-$super-space
without torsion may correspond to a space, for which
$y^{\mu}y_{\mu}=0$ only.
\par
For the substitution of the spinor coordinates $\chi_{a}$ for the
vectorial ones $y^{\mu}$ we use an equation of the type (2)
\begin{equation}
y^{\mu}y_{\mu}=s^{2}
\end{equation}
and its linearized form:
\begin{equation}
\gamma^{\mu}y_{\mu}\chi=s\chi,
\end{equation}
where $\gamma^{\mu}$ are the Dirac matrices, $\chi$ is a bispinor,
$y_{\mu}$ and $s$ are the same as in (34) - they are not
operators, but usual numbers. From (35) one obtains the relation
of connection of vectorial $y_{\mu}$ and spinor
$\chi_{a},\bar{\chi}_{a}$ coordinates, $a=1,2,3,4:$
\begin{equation}
y^{\mu}=\bar{\chi}\gamma^{\mu}\chi,
\end{equation}
and (34) is carried out by some additional restrictions on
$\chi_{a},\bar{\chi}_{a}.$ If all $\chi_{a},\bar{\chi}_{a}$
commute, it is necessary (compare with \cite{31}) that
\begin{equation}
 R^{2}=0,
\end{equation}
where $R=\bar{\chi}\gamma^{5}\chi,\
\gamma^{5}=i\gamma^{0}\gamma^{1}\gamma^{2}\gamma^{3},$ and, if
these anticommute, the condition must be carried out
\begin{equation}
R^{2}+8M =0,
\end{equation}
where
$M=\bar{\chi}_{1}\chi_{1}\bar{\chi}_{4}\chi_{4}+(\bar{\chi}_{2}\chi_{2}+
\bar{\chi}_{1}\chi_{1})\bar{\chi}_{3}\chi_{3}
+\bar{\chi}_{1}\chi_{4}\bar{\chi}_{2}\chi_{3}+
\bar{\chi}_{3}\chi_{2}\bar{\chi}_{4}\chi_{1}.$ For this the
dependence $\bar{\chi}=\chi\dagger\gamma^{0},\
\chi_{a}^{*}=\chi_{a},$ leads to identity in case of (37) and to a
local variant of the composite fermions model in the case of (38).
\par
One must consider $\chi_{a},\bar{\chi}_{a},$ to be independent of
$x^{\mu},$ taking into account the postulate c), and for this
$\chi_{a},\bar{\chi}_{a}$ may commute or anticommute between
themselves \cite{15}. In last case, $(x,\bar{\chi},\chi)$ is the
$(N=2)-$superspace without torsion. The substitution $\gamma^{\mu}
\rightarrow \gamma^{\mu}_{A}$ in (36) permits to get the set of
$y^{\mu}_{A},$ if $\gamma^{\mu}_{A}=\gamma^{\mu}\Gamma_{A}
B^{\mu}_{A},$ where $B^{\mu}_{A} \in \{\pm I^{4},\pm \gamma^{5}
\},\ B^{\mu}_{1}=I^{4} \ \forall \mu,\ \Gamma_{1}=I^{4},\
\Gamma_{2}=iI^{2}\times \sigma^{1},\ \Gamma_{3}=\sigma^{1}\times
\sigma^{2},\ \Gamma_{4}=\sigma^{1}\times \sigma^{3},\
\gamma^{\mu}_{A+4}=-\gamma^{\mu}_{A}.$ Matrices $B^{\mu}_{A}=\pm
\gamma^{5}$ only for $y^{\mu}_{A}=0,$ so that in the quark sector
of the model
\begin{equation}
y^{\mu}_{A}=0\ \Rightarrow \bar{\chi} \gamma^{\mu}
\gamma^{5}\chi_{A}=0,\ y^{\mu}_{A}=\pm l \ \Rightarrow \bar{\chi}
\gamma^{\mu} \chi_{A}=\pm l,
\end{equation}
where $\chi_{A}=\Gamma_{A}\chi$ and in the last line the inverse
correspondence of signs is also possible.
\par
Thus, the $(N=2)-$superspace without torsion $(x,\bar{\chi},\chi)$
with the set of ties
\begin{equation}
y^{\mu}_{A}=\bar{\chi}\gamma^{\mu}_{A}\chi=
Re{\bar{\chi}\gamma^{\mu}_{A}\chi},
\end{equation}
picking out $8$ supermanifolds, may be represented on the space
$Z=(x,y^{\mu}_{A})$ of the model: $x \rightarrow  x,\
(\bar{\chi},\chi) \rightarrow y^{\mu}_{A}$ by formulae (40). Then
one may consider an internal symmetry as a consequence of mixing
of solutions, which are defined on different manifolds. The
chirality of $SU(2)$ is caused by the field equations as before,
but it is not connected with any peculiarities of a basic space
$(x,\bar{\chi},\chi).$
\par
Rejection of the axioms of the supersymmetrical models about the
presence of torsion and about the dependent transformation of $x$
and $\bar{\chi},\chi$ changes essentially an interpretation of
component fields, which are contained by any scalar superfield in
a space $(x,\bar{\chi},\chi)$ - now all of them must be scalar
with respect to the Lorentz group of a basic space $X.$ The
Lagrangian of the model, being considered as such superfield, must
contain terms of expansion on degrees of $\bar{\chi},\chi;$
factors by the first degree of $\chi_{A}$ must be the sets of
scalars with respect to the Lorentz group and $SU(2)-$doublets,
i.e. these are analogs of the Higgs fields of the standard model.

\section[10]{Conclusion}

The interpretation of the composite fermion model, given here, is
based essentially on investigations of local symmetries and gauge
interactions, which have been made in the frame of the standard
model. One may say that an assumption about the composite nature
of the fundamental fermions permits not only to describe the
physics of particles, as the standard model does with success, but
and to explain this. In the composite fermions model, the basic
propositions of the standard model go into the common logically
harmonious picture, their ties and an interconditionality are
visible. In such a context the standard model may be interpreted
as a way to describe non-local interactions of composite objects,
and its experimental corroboration may be interpreted as an
indirect confirmation of the composite nature of the fundamental
fermions.
\par
In the text, the postulate of the standard model about mixing of
generations in the quark sector was not mentioned. A similar
phenomenon takes place in the composite fermions model, but to
evaluate this effect, its description in the model, based on
introduction of $4$ generations at once, will be be made to agree
with the conception of mixing in the standard model language. The
representation for matrices $G_{A}$ \cite{14} is such that
matrices $G_{6}$ and $G_{7}$  mix generations in pairs in the
quark sector - in the sense that if, for instance, in the lepton
sector some wave function components set up the first generation,
then in the quark one a mixture of components of the first
generation and of the third one corresponds to this (here the
numbering of the generations corresponds only to the indices of
$\psi$) \footnote{for another pair of generations, one has the
same effect in the quark sector of the model}.
\par
\vspace{0.3cm} {\bf PS to this reprint}\\ It is a reprint of
author's paper of 1992. I am grateful to the editors of Nuovo
Cimento for a permission to re-print it. I think that after the
observation of evidence of non-zero neutrino mass difference by
the Super-Kamiokande collaboration \cite{100}, my study may be
more actual than ten years ago. Another reason to enter this paper
into the LANL archive is my recent study of the hypothetical
background of super-strong interacting gravitons
\cite{101,102,103}. Perhaps, such gravitons would be very
appropriate candidates to a role of constituents of the composite
fundamental fermions.

%\begin{references}

%\end{references}
\end{document}